\documentclass{article}

\usepackage[english]{babel}

\usepackage[letterpaper,top=2cm,bottom=2cm,left=3cm,right=3cm,marginparwidth=1.75cm]{geometry}

\usepackage{graphicx}	% Including figure files
\usepackage{amsmath}	% Advanced maths commands
\usepackage{bm}
\usepackage{algorithm}
\usepackage{physics}
\usepackage{caption}
\usepackage{subcaption}
\usepackage{natbib}
\usepackage{hyperref}
\usepackage{xcolor}
\usepackage[normalem]{ulem}
\usepackage{placeins} % FloatBarrier and stuff
\usepackage[displaymath, mathlines]{lineno}
\title{Estimating noise for airborne electromagnetic data from repeat flight lines or inversion residuals}
\date{}
\newcommand*\samethanks[1][\value{footnote}]{\footnotemark[#1]}
\author{Tim Scarr\thanks{Geoscience Australia, email:tim.scarr@ga.gov.au} \and Anandaroop Ray\samethanks
\and Ross C. Brodie\thanks{Formerly at Geoscience Australia}}
\begin{document}
\label{firstpage}
\maketitle

\begin{abstract}
Characterising the noise of an airborne electromagnetic (AEM) system is critical in correctly imaging the earth's subsurface conductivity. Deterministic and probabilistic geophysical inversion algorithms require foreknowledge of the system noise to specify stopping criteria or a valid model likelihood. Repeat flight lines provide a way for geophysicists to calculate the statistical variability in AEM data acquired over the same ground, and therefore estimate the levels of noise to propagate into the inversion. The total noise can be separated into multiplicative and additive components. The multiplicative noise is derived by repeat lines at survey altitude. The method to calculate the multiplicative noise is scarcely documented and usual methods for height correcting acquired data require a linear trend removal. This study will outline the algorithm used to estimate multiplicative noise of an AEM system, and non-linearly correct for varying altitudes during repeat flights. Additionally, this paper details a methodology to Gaussianise the data noise and provide a statistically valid Gaussian data misfit or likelihood function. Significantly, we provide methods for estimating the off-diagonal elements in the data covariance matrix used within the misfit function, taking into account the time-channel data correlation that is usually neglected. While our methodology is general, our study of a rotary-wing system leads us to conclude that for regularised time-domain AEM imaging, a diagonal data covariance suffices -- an important implication for rigorous yet practical AEM inversion.
\end{abstract}

% Include between one and six keywords.
Data Methods -- Instrumentation -- Electromagnetics -- Inversion -- Noise Statistics -- Airborne

%%%%%%%%%%%%%%%%%%%%%%%%%%%%%%%%%%%%%%%%%%%%%%%%%%

%%%%%%%%%%%%%%%%% BODY OF PAPER %%%%%%%%%%%%%%%%%%

\section{Introduction}
\label{sec:intro}
\begin{figure*}
    \centering
    \includegraphics[width= 0.75\linewidth]{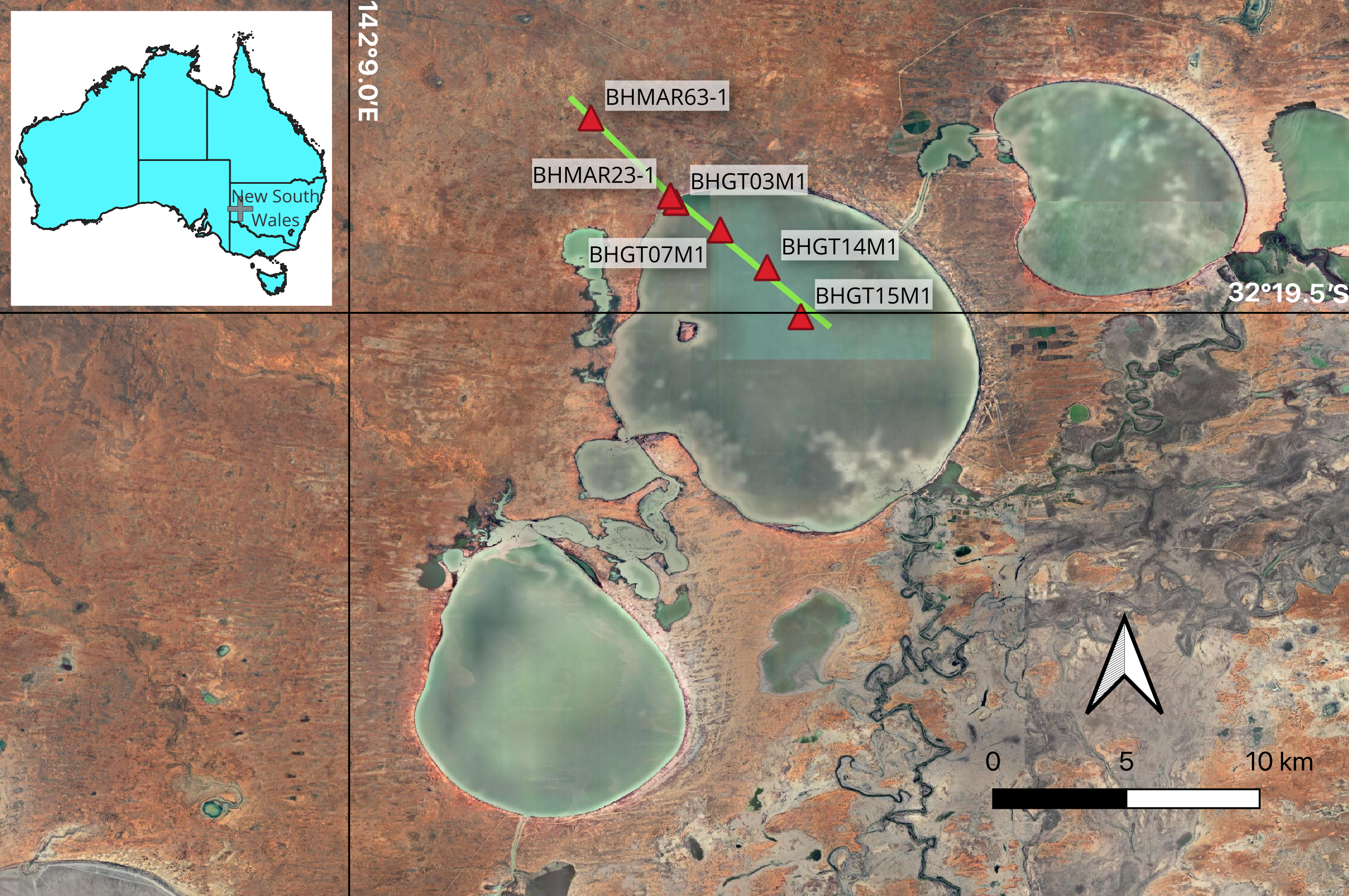}
    \caption{Repeat line section over Menindee Lakes, New South Wales, Australia. The green line shows the prescribed repeat line path, flown by NRG with the 25Hz XCITE rotary-wing system in 2024. Red triangles are boreholes with associated names.}
    \label{fig:menindee}
\end{figure*}
Airborne electromagnetic (AEM) prospecting is a useful geophysical technique for imaging shallow subsurface features when conducting large-scale land-based surveys. AEM data are collected by pulsing the earth with an electromagnetic field. In accordance with Maxwell's equations and particularly Faraday's Law of electromagnetic induction, these pulses elicit electromagnetic responses from conducting layers in the subsurface, down to approximately 300 m depth \citep{vozoff_1980, brodie_2010}. Using AEM to understand the electromagnetic properties of the subsurface has significant applications in mineral exploration and geological mapping \citep{murr_2020, wong_2020}, groundwater detection \citep{kirsch_2006, minsley_2021, kang_2022}, and subsurface mapping for national scale surveys \citep{leycooper_2020}. AEM can advantageously be deployed over various terrains with limited accessibility and allows for fast data collection (100-300 km / hr) at line spacings from 100 m to 20 km \citep{wu_2024}.

To obtain the conductivity profile of the subsurface from AEM data, we use inversion theory to calculate the geoelectric properties from voltage or magnetic field that is acquired at the AEM receiver \citep{parker_1994, oldenburg_1999, brodie_2010, menke_2012, zhadanov_2015}. This is a non-linear, non-unique problem that requires careful consideration of model constraints and data uncertainties to produce a useful and interpretable conductivity model. There are two main AEM inversion approaches that are practised. Deterministic inversion converges to an optimal solution by iteratively updating the model to minimise the data misfit, while honouring the constraints of the model \citep{constable_1987, vallee_2009, brodie_2016, auken_2015, key_2016}. Probabilistic inversion takes a stochastic approach that treats the earth conductivity model as a probability distribution. Through Bayesian inference, prior knowledge and data uncertainties are used for posterior sampling of this distribution using approaches such as Markov chain Monte Carlo \citep{moosegard_1995, malinverno_2002, sambridge_2002, tarantola_2005, blatter_2018, ray_2023}. Probabilistic inversions provide non-linear model uncertainty, whereas deterministic inversions provide a single, smooth model at far lower computational expense \citep{valentine_2019}. In this study we focus on deterministic inversion since our focus is on characterising data uncertainty, as opposed to model uncertainty.  However the main conclusions drawn are equally applicable to Bayesian inversion. 

There are multiple causes for noise in AEM data and are categorised as being either systematic or random \citep{brodie_2010, davis_2023}. Systematic noise is from consistent and known patterns of interference from sources such as imprecise primary field cancellation or estimation, sferics (i.e., lightning) and power lines \citep{brodie_2010}. These can be corrected by filters, synchronous stacking or knowing the source. Random noise is caused by events with random, statistical variation of the signal received. 
Both deterministic and probabilistic inversion use a misfit (or negative log likelihood) function that is minimised. Use of Gaussian likelihood or data misfit functions are justified by the data stacking process that increases signal-to-noise through variance reduction, in accordance with the Central Limit Theorem. The likelihood also includes a data covariance matrix that captures this statistical noise, which propagates to the inverted conductivity model \citep{constable_1987, agnostinetti_2010}. Appropriately formulated noise models, misfit functions and inversion schemes should honour the data noise statistics \citep{dosso_2006}. The aforementioned statement is scrutinised closely and appropriately qualified in this work, as deterministic inverse model estimates invariably lead to correlated inversion residuals. 

The most widely-used noise model for AEM was formulated by \citet{green_and_lane}, who propose that the dominant source of noise at survey altitude is proportional to signal amplitude at early to mid times, and dominated by additive noise at late times. Many algorithms and methods exist to remove noise and to minimise its effect on the data \citep{peng_2020, wu_2020}. However the extent to which noise is ``removed'' should be treated with caution \citep{mule_2019}, as this may impact the noise statistics \citep{Dettmer2012a} and ultimately, the inverted conductivities. Further, ignoring off-diagonal elements in the data covariance, as is current usual practice due to limited capacity to estimate off-diagonal terms, can lead to artefacts as demonstrated by various workers in geoacoustic, seismic and controlled source electromagnetic problems \citep[e.g.,][]{dosso_2006, bodin_2012, ray_2013}.

The novelty of our work is manifold. First, we detail an inversion based, non-linear height correction for the AEM data that allows us to derive the multiplicative noise model for all channels. A variant of this method has been used internally at Geoscience Australia since the publication of a report on the Southern Thomsons Orogen Project \citep{roach_2015}, but is not in the peer reviewed literature. Next, we find a ``most-Gaussian" data variance by forcing the multiplicative noise per time channel to be as closely distributed to a Gaussian cumulative distribution function (CDF) as possible through a non-linear optimisation process. We then estimate and use within our inversions, the off-diagonal covariance elements either from repeat lines, or inversion residuals following the method of \citet{dosso_2006} that does not require repeat lines. 
\section{Theory}
\subsection{AEM data noise model}

As is usual practice for AEM data $\mathbf{d}$ with linear zero mean Gaussian noise $\mathbf n$ with covariance $\mathbf{C_d}$, we use the noise model:
\begin{equation}
	\mathbf{d} = \mathbf{f(m)} + \mathbf{n}; \;\text{where }\mathbf{n} \sim \mathcal{N}\mathbf{(0, C_d)}.
\end{equation}
We use the AEM noise model first proposed by \citet{green_and_lane} as follows. High altitude flights are used to estimate the standard deviation of the additive noise, and repeat lines are used to estimate the multiplicative component. These two components are assumed to be independent, so the noise standard deviation $\sigma_j$ for channel $j$, is defined as follows:
\begin{equation}
    \label{eq:noise_equation}
    \sigma_j = \sqrt{(k d_j)^2 + {a_j}^2},
\end{equation}
where $d_j$ is the $j$-th time Z-component data and $a_j$ is the standard deviation of additive noise in the $j$-th channel. $k$ is the multiplicative noise factor such that $kd_j$ is the standard deviation of multiplicative noise in the $j$-th channel. The words ``channel'' and ``window'' are commonplace in AEM parlance and used interchangeably in this work. In the presence of appreciable earth response and early time channels, the multiplicative noise dominates, whereas for low-signal resistive terranes or late time channels, the additive noise dominates.

Focussing on the dominant noise source in the high signal amplitude regime, we can better understand the reasoning behind a multiplicative noise model as follows. Diffusive electromagnetic data span many orders of magnitude, and are typically observed to be logarithmically scaled, and authors such as \cite{Wheelock2015} have explicitly made this claim and studied error propagation in this light. Though we do not take the logarithm of the data in our work, the following analysis is instructive. If we consider some observable with logarithmic scaling $x_o$ contaminated by Gaussian noise $\epsilon$ such that
\begin{align}
\log x_o &= \log x + \epsilon; \;\text{where }\epsilon \sim \mathcal{N}(0, \sigma^2)\\
&= \log x + \log e^\epsilon,\\
x_o &= xe^\epsilon. \\
\intertext{For $\sigma \ll 1$, $\epsilon$ has high probability of being close to zero,}
x_o &\approx x(1+\epsilon) \;\text{to first order},\\
\implies \frac {x_o -x}{x}&\sim \epsilon.
\intertext{For small noise such that $x$ and $x_o$ are close, this further implies that the standardised variable}
\frac {x_o -x}{\sigma x_o}&\sim \mathcal{N}(0, 1) \;\text{to first order}. \label{eqn:multnoise_deriv}
\end{align}  
Motivated by this reasoning we continue our analysis with the explicit assumption that it is the factor $k$ in \eqref{eq:noise_equation} analogous to $\sigma$ in \eqref{eqn:multnoise_deriv} that we are estimating in this work. The assumed noise model in \eqref{eq:noise_equation} is common in AEM applications \citep[e.g.,][]{Minsley2021a}, with the additive noise allowing for its use in both high and low signal regimes.

\begin{figure}
    \includegraphics[width=\columnwidth]{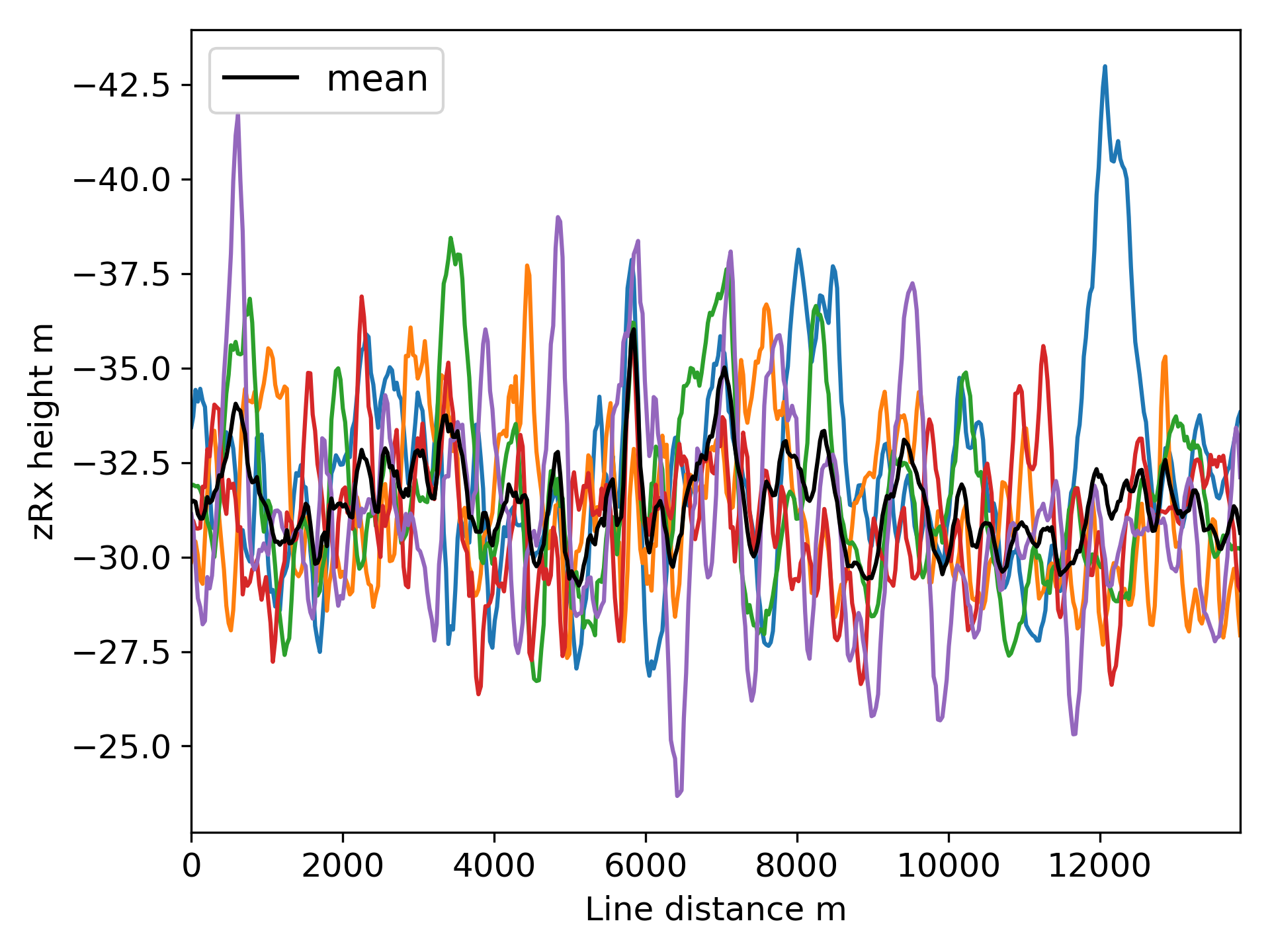}
    \centering
    \caption{Plot of receiver heights and average of AEM system across all repeat lines flown over Lake Menindee. Different colours represent different repeat lines. z is positive downwards into the earth. The thick, black line highlights the mean height from all lines.}
    \label{fig:flight_heights}
\end{figure}
\begin{figure}
    \includegraphics[width=\columnwidth]{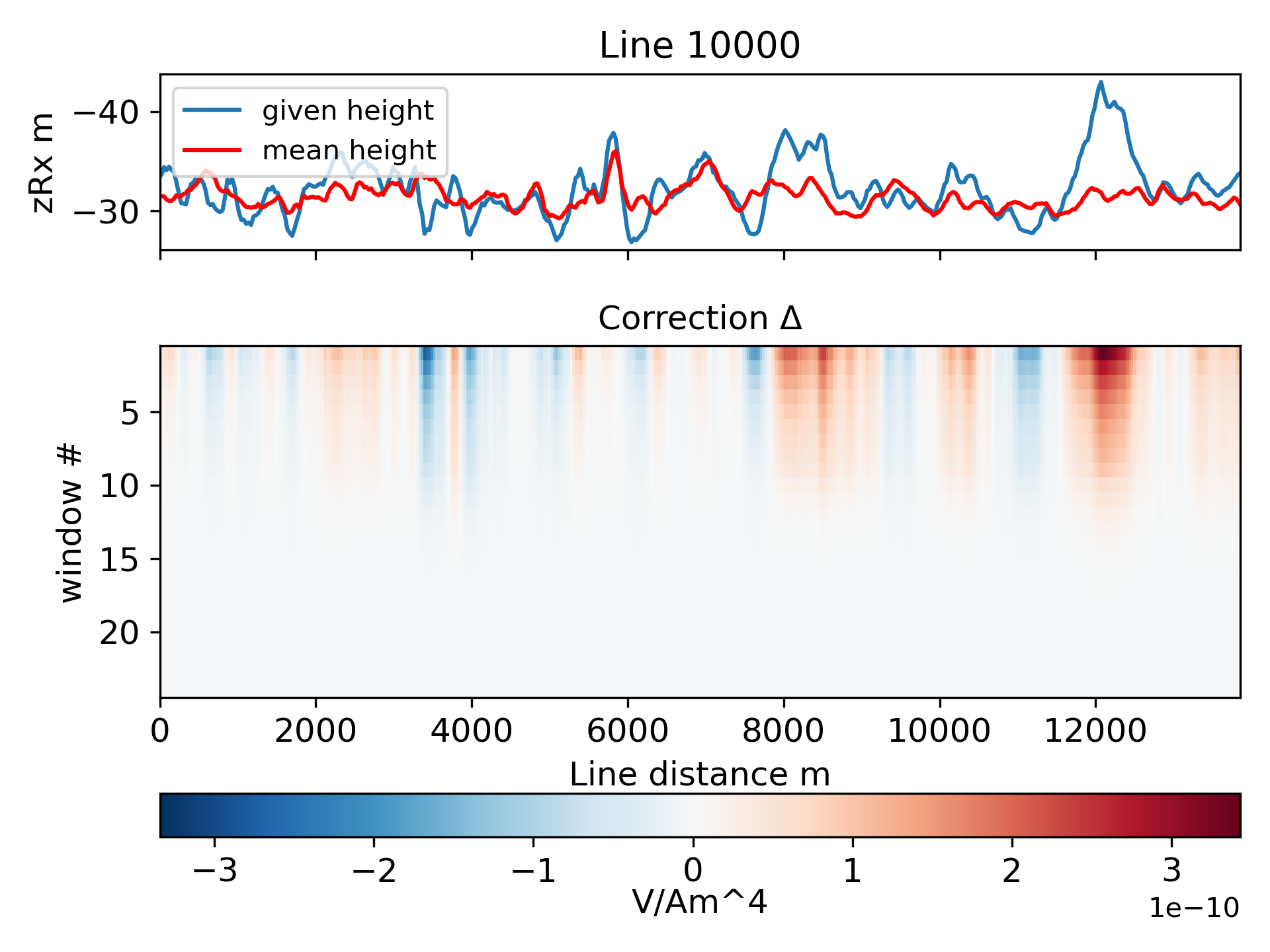}
    \centering
    \caption{Plot of corrections $\Delta_j$ for the differing receiver heights across all time channels (window \#). Note how when the actual, acquired receiver height is above the desired height (given by the mean from all flights), we need to add to the amplitude of the signal (z is positive downwards into the earth). This is because the secondary field amplitude at higher (than desired) height is lower, and requires positive residuals to be added to the field to correct it. Similarly, $\Delta_j$ is negative when the actual receiver height is below the desired observation height. The correction is strongest in the early time channels.}
    \label{fig:corrections}
\end{figure}
\subsection{Deterministic inversion framework}
In this paper we used the Occam inversion framework to perform deterministic inversion on AEM data \citep{constable_1987, key_2016}. The algorithm aims to minimise the following objective function:
\begin{equation}
\label{eq:deterministic_misfit}
    \phi(\textbf{m}) = \norm{\textbf{W}(\textbf{d}-\textbf{f}(\textbf{m}))}^2 + \lambda^2 [\norm{\textbf{Dm}}^2 + \beta^2 \norm{\textbf{m} - \mathbf{m_0}}^2],
\end{equation}
where \textbf{m} is the conductivity model, \textbf{f} is the AEM forward operator containing data \textbf{d} (voltage), and \textbf{W} are the data weights that relate to the data covariance matrix $\mathbf{C_d}$ such that $\mathbf{C_d^{-1}} = \mathbf{W^tW}$. \textbf{D} is a first differences regularisation operator, $\lambda^2 >0$ is the regularisation parameter, $\mathbf{m_0}$ is a reference model, and $\beta^2$ is a positive scalar that dictates bias towards the reference model. Based on the Occam's inversion algorithm, we linearise the problem by adopting a Gauss-Newton approach to minimise the objective function. This involves sweeping through $\lambda^2$ to identify the largest (and therefore smoothest) $\lambda^2$ value that fits the model within the accepted data noise. 

To show the connection with Gaussian data errors and their $\chi_n^2$ distribution, we first write the normalised data misfit $\phi_d$ as follows:
\begin{align}
\phi_d (\mathbf W, \mathbf m) &= \frac 1 n {\chi_n^2} (\mathbf W, \mathbf m) = \frac 1 n \mathbf{||W(d - f(m))||}^2, \label{eq:phid}
\intertext{keeping in mind that} 
\mathbf{C_d^{-1}} &= \mathbf{W}^t\mathbf{W}. \label{eq:wtw}
\end{align}
where $n$ are the number of data within a sounding. Finally, to write the above equation in terms of a likelihood function that provides the probability of the observations given the model or $p\mathbf{(d|m)}$, we know from the Gaussian probability density function that
\begin{equation}
p\mathbf{(d|m)} \propto \exp[-\frac n 2 \phi_d (\mathbf W, \mathbf m)], \label{eq:gauss}
\end{equation}
ignoring constants that are not dependent on the inversion model. As is well known for $\chi_n^2$ errors, their expected value is $n$ \citep{parker_1994}, thus making 1 the target $\phi_d$ value for Occam's inversion.
\subsection{Correlation of data errors}
\label{sec:correlation}
Standard approaches to AEM inversion assume data errors are Gaussian distributed and uncorrelated between channels. In practice, noise and data residuals have the potential to be correlated across time channels. This may be from uncertainty in transmitter position, theory error, model parameterisation choices, low pass filtering as well as other factors listed in \cite{brodie_2010}. Efforts in probabilistic geophysical inversion to address this have been found in geoacoustics, surface wave dispersion and marine controlled source electromagnetics \citep{dosso_2006, bodin_2012, ray_2013}. For data transformations (which naturally transform the data error), a deterministic inversion example with non-diagonal covariance for the magnetotelluric problem can be found in \cite{moorkampUsingNondiagonalData2020}. The misfit function in these cases include a data covariance matrix $\mathbf{C_d}$ as described in equations~\ref{eq:phid} and~\ref{eq:wtw}. $\mathbf{C_d}$ can be broken into a diagonal matrix of standard deviations \textbf{S} and a correlation matrix \textbf{R}, such that 
\begin{equation}
    \mathbf{C_d} = \mathbf{SRS}, \label{eq:Cd_decomposition}
\end{equation} 
where each element of the correlation matrix, $r_{ij}$, represents the Pearson correlation coefficient \citep{profillidis_2019} between two time channels $i$ and $j$. If post inversion residuals $\mathbf{r}$ are distributed according to the covariance specified in the likelihood function \eqref{eq:gauss}, then $\mathbf{C_d}$ can be decomposed with an upper Cholesky factor $\mathbf U$ such that
\begin{align}
\mathbf{C_d} &= \mathbf{U}^t\mathbf{U}. \label{eq:utu}
\intertext {From a theoretical perspective, perfectly whitened residuals $\mathbf{r_w}$ can be obtained by premultiplying $\mathbf{r}$ with the inverse transposed Cholesky factor as follows:}
 \mathbf r_w &=  \mathbf U^{-t}  \mathbf r.
 \intertext{This can be seen by obtaining the whitened covariance of zero mean residuals}
   \big< \mathbf r_w  \mathbf r_w^t \big> &=  \mathbf U^{-t}    \big< \mathbf r \mathbf r^t    \big>  \mathbf U^{-1} =\mathbf U^{-t}   \mathbf{C_d}   \mathbf U^{-1}, \\
   & = \mathbf U^{-t}   \mathbf U^t \mathbf U  \mathbf U^{-1}  = \mathbf I,
\end{align}
where $\big<\cdot \big>$ is the expectation operator and $\mathbf I$ is the identity operator.

\section{Methodology for noise estimation from repeat lines}
\subsection{Calculating $k$ as a constant for all times}
\label{sec:repeat_lines_estimating_method}
Repeat lines in a survey are used as a quality control measure, as well as for a statistical estimation of multiplicative noise, based on the repeatability of data measured on multiple occasions over the same ground. While repeat lines aim to survey the same line, the receiver height of the system can change with each repeat flight, which significantly impacts the recorded earth response. The original  \citet{green_and_lane} paper offered a linear, regression based height correction which we improve upon. An earth model based non-linear height correction can be applied to the data, which we outline here with five repeat lines flown in 2024 with the New Resolution Geophysics (NRG) 25Hz XCITE rotary-wing system over Menindee Lakes, New South Wales (Figure \ref{fig:menindee}). According to the noise estimation formula \eqref{eq:noise_equation}, estimating the multiplicative noise requires the AEM response data $\mathbf{d}$, as well as a factor $k$. 

The algorithm begins with a first guess inversion, with $k = 0.05$ on all repeat line passes, to generate as many conductivity models as there are repeats. As the first pass estimate of $k$ is relatively high, each inversion is guaranteed to fit a smooth conductivity model to the data. The first guess inverse models and associated soundings are resampled at X and Y node locations on the earth (using a nearest neighbours technique) on a single, common line. This common line is found from a smooth least squares fit for a line that goes between the various repeat line paths. A mean conductivity-depth model at each sounding is then calculated from the various inversion models, assumed to be representative of the earth conductivity $\mathbf{m_e}$ underneath the line. We designate the mean height of the system receiver at each node as $z_\text{wanted}$.  Each repeat line sounding is then forward modelled, using the mean conductivity model $\mathbf{m_e}$ at both $z_\text{wanted}$, as well as at the actual recorded height $z_\text{acquired}$ (Figure~\ref{fig:flight_heights}).

\begin{figure}
\centering
    \includegraphics[width = \columnwidth]{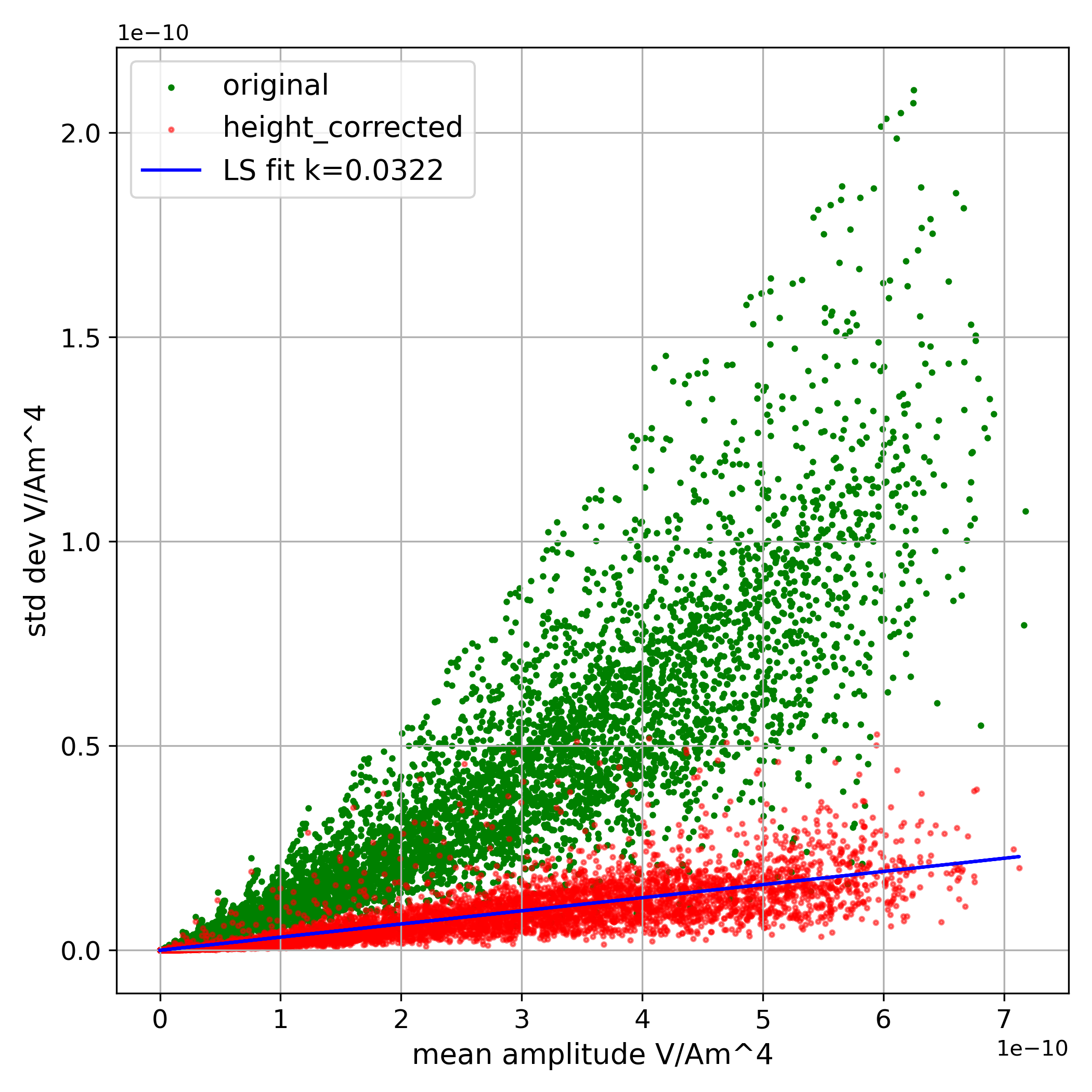}
    \caption{Comparison cross-plot of standard deviation vs mean of Z-component AEM data. Green dots show the non-height corrected data and the red dots show the height corrected data. The blue line represents the least squares fit of the height corrected data, the slope of which, $k$, is the multiplicative noise factor for the system.}
    \label{fig:mean_vs_std}
\end{figure}

Assuming the earth conductivity $\mathbf{m_e}$ is representative of the earth, the difference between the forward calculations at $z_\text{wanted}$ and $z_\text{acquired}$ is treated as the height corrected (HC) data residual $\Delta$. Examples of this correction for height can be seen along a particular repeat line in Figure~\ref{fig:corrections}. The HC data $d_\text{wanted}$ for each line at each sounding and channel is then calculated by adding $\Delta$ to the observed data $d_\text{acquired}$. Subsequently, the mean of the HC data $d_\text{wanted}$ and its standard deviation is calculated at each time channel for each node along the line. The standard deviation vs mean data amplitude from all nodes across all channels can be examined as a linear model where the slope of the least squares fit is the multiplicative noise estimate of $k$ (Figure~\ref{fig:mean_vs_std}). This highlights the success of the non-linear height correction on the apparent line-to-line repeatability since the slope and spread of the corrected (red) dots is far less than the uncorrected (green) dots. See algorithm \ref{alg:mult_noise} for a methodical outline of the steps to calculate $k$. Note that this estimate is the same for all time channels. It is remarkable that there are striking similarities between our airborne approach and that of \cite{gehrmannSeafloorMassiveSulphide2019} for navigational errors in marine electromagnetic surveying, though their uncertainty analysis is perturbative and our height corrections are not.

\begin{algorithm}
    \footnotesize
    \begin{enumerate}
    \item{ First-guess deterministic inversion using a high guess of $k = 0.05$ on all passes of the repeat lines.}
    \item{Find a common smooth line path in a least squares sense, between all repeat paths.} 
    \item{Use nearest neighbours interpolation to find the nearest sounding data and inverted conductivity model from the repeat lines onto this line.} 
    \item{Construct a mean model $\mathbf{m_e}$ at all locations along the common line path.}
    \item{From all the repeat lines, calculate the mean transmitter height $z_\text{wanted}$ at each line location.}
    \item{For each sounding at all time channels $j$, forward calculate the residual $\Delta_j=f_j\mathbf{(m_e}, z_\text{wanted}) - f_j\mathbf{(m_e}, z_\text{acquired})$}	
    \item{Assume that recorded heights are perfectly known, and so $d_j(z_\text{wanted}) = d_j(z_\text{acquired}) + \Delta_j$ 
    \item{Plot the standard deviation \textit{vs} the mean of $d_j(z_\text{wanted})$ at all common line locations, for all time channels $j$ \label{item:corrected_data}(Figure~\ref{fig:mean_vs_std}).}
    \item{The slope of the least squares fit in Figure~\ref{fig:mean_vs_std} provides the estimate of multiplicative noise factor $k$ to use in \eqref{eq:noise_equation}.}
}
\end{enumerate}
\caption{Multiplicative noise factor estimation}
\label{alg:mult_noise}
\end{algorithm}
\begin{figure}
\centering
    \includegraphics[width=\columnwidth]{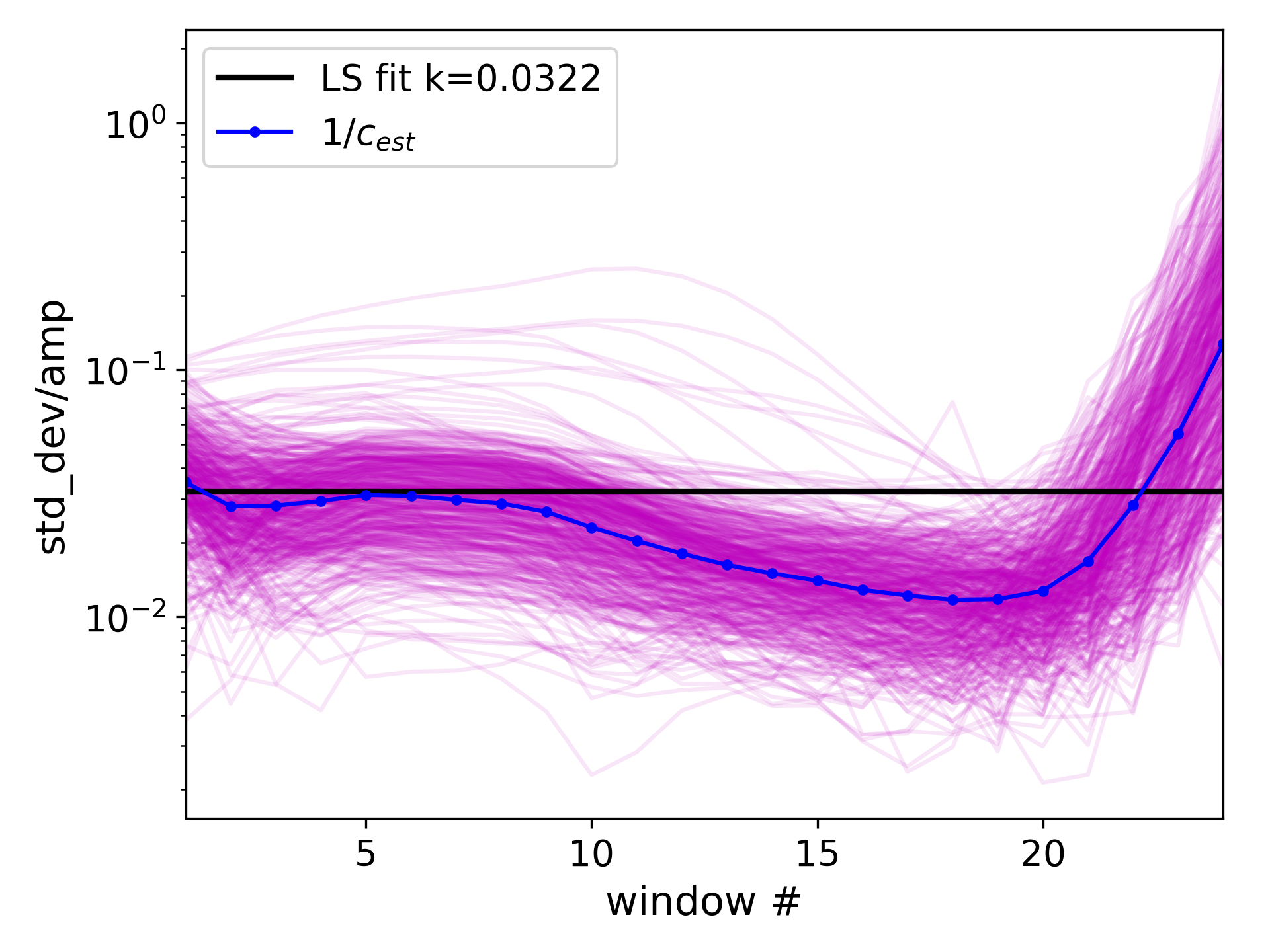}
    \caption{Plot of standard deviation divided by mean amplitude of the HC data, plotted against time channel (window \#), for all soundings from all repeat lines. Black straight line represents the least-squares fit (Figure~\ref{fig:mean_vs_std}) for the multiplicative noise factor calculated from the method in section \ref{sec:repeat_lines_estimating_method}. Blue line shows the optimised multiplicative noise per channel from Section~\ref{sec:noise_vec}.}
    \label{fig:noise_vec}
\end{figure}
\begin{figure}
        \includegraphics[width=\columnwidth]{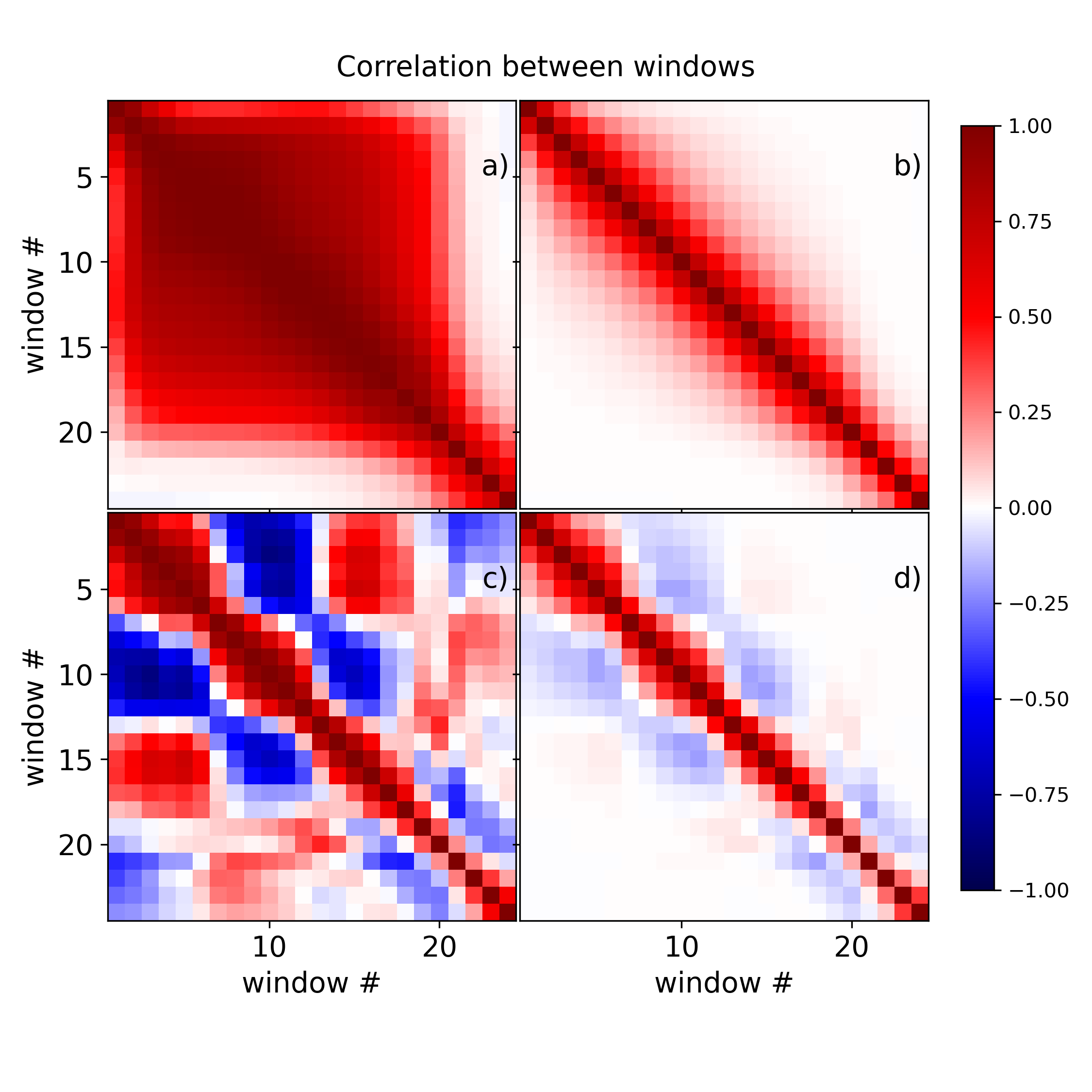}
         \caption{Correlation matrix from height corrected repeat line data (a,b) or inversion residuals (c,d), varied by roll-off function. (a,c) are undamped while (b,d) use exponential roll-off.}
    \label{fig:correlations}   
\end{figure}
\subsection{Calculating $k$ per time channel}
\label{sec:noise_vec}
In our experience, $k\approx0.03$ is a reasonable estimate for the multiplicative noise for helicopter systems flown with receivers and transmitters between 30-40 m above ground (Figure~\ref{fig:mean_vs_std}). However, to additionally satisfy the assumptions of Gaussianity in~\eqref{eq:phid} or~\eqref{eq:gauss}, we examine the effect of fitting the height corrected deviations using a Gaussian cumulative distribution function (CDF). This method essentially scales the unit Gaussian until its CDF matches the height corrected observations. Using the height corrected data, given a CDF $x$ of data deviations from the mean, divided by mean data in a channel, we find the factor $c$ by minimising the following objective function using gradient descent: %
\begin{equation}
\label{eq:noise_vec}
    \Phi = \sum_i^{\substack{\text{all} \\ \text{soundings}}}\Big[\text{CDF}(x_i)-\text{CDF}_{\mathcal{N}(0,1)}(cx_i)\Big]^2.
\end{equation}
The factor $\frac 1 c$ per time channel, is the time-varying multiplicative noise estimate $k_j$, henceforth referred to as the ``vector noise" estimate (Figure~\ref{fig:noise_vec}). The rationale behind this objective function is explained in Appendix~\ref{sec:rationale}.
\subsection{Estimating inter-channel correlation}
\label{sec:iccorr}
There are two methods that can be used to calculate correlation. We can 1) use the repeat line data and their statistics, or 2) the approach from \citet{dosso_2006} that uses inversion residuals in a geoacoustic inverse problem. \cite{Scales1998} use a similar strategy to estimate correlation for a full waveform seismic inversion problem. Both follow the same foundational principle, starting from equation~\eqref{eq:Cd_decomposition}.

The residuals, either from height corrected repeat line data, or from inversion residuals are denoted as a random vector $\mathbf x$, representative of the noise process in the data. We can then write, using equation~\eqref{eq:Cd_decomposition} that:
\begin{align}
\mathbf x &\sim \mathcal N(\mathbf{0, SRS}).
\end{align}
When using repeat lines, $\mathbf S$ can be estimated from the methods of Sections~\ref{sec:repeat_lines_estimating_method} or \ref{sec:noise_vec}. The correlation $\mathbf R$ is obtained from the height corrected data -- one realisation of residuals at a sounding location is calculated by subtracting the height corrected data from the mean of the height corrected data. As there are too few repeat lines and many more time channels, we cannot robustly calculate a sounding-by-sounding correlation matrix. This is why, similar to the case of estimating multiplicative noise (Step~\ref{item:corrected_data} in Algorithm~\ref{alg:mult_noise}), we assume that the noise process is the same across the survey, and concatenate all such residuals to obtain $n_\text{soundings}$ realisations of $\mathbf x$ to calculate correlation $\mathbf R$ from.

For the case of first-pass inversion residuals $\mathbf{x = f(m) - d}$ standardised only by a diagonal $\mathbf S$, let us denote another random vector:
\begin{align}
\mathbf x_* &\equiv \mathbf S^{-1}\mathbf{x}.
\intertext{The elements of this diagonally whitened random vector $\mathbf x_*$ have the following covariance:}
\mathbf{C}_\text{est} &= \big<\mathbf x_* \mathbf x_*^t\big>,\\
&=\big<\mathbf S^{-1} \mathbf {xx}^t\mathbf S^{-1}\big>,\\
&=\mathbf S^{-1} \big<\mathbf {xx}^t \big> \mathbf S^{-1},\\ 
&= \mathbf S^{-1} \mathbf{SRS} \mathbf S^{-1},\\
&= \mathbf R.
\end{align}

Further, \citet{dosso_2006} suggests that elements far from the diagonal of a correlation matrix $\mathbf{R}$ should be small, can often be poorly estimated, leading to un-physical correlation estimates. Ultimately, a damping function can be included in the estimated correlation. This brings to zero, elements in the correlation matrix that are far from the main diagonal. The damping function used in this paper is:
\begin{align}
%   s_{i,j} &= cos^l\frac{\pi\abs{i-j}}{2(n-1)}, \;i,j = 1,...,n, \label{eq:cos_roll-off}\\
%\nonumber \text{and,}\\ 
    s_{i,j} &= e^{-l |i-j|}, \;i,j = 1,...,n, \label{eq:exproll-off}
\end{align}
where $l$ determines the drop-off rate which is selected to approximately approach zero by $\frac n 2$ elements away from the diagonal (Figure~\ref{fig:correlations}). Such a damped correlation matrix is then plugged back into the data misfit function in Equation~\ref{eq:phid} for an inversion that utilises correlation, noting the connection between $\mathbf W$ and $\mathbf {C_d}$ in Equations~\ref{eq:wtw} and \ref{eq:Cd_decomposition}. 

At this stage, we must mention that there are important caveats on obtaining correlation either involving, or from a least squares inversion estimate. This is discussed later in Section~\ref{sec:normalised_residuals_analysis}, which lead to significant departures from the Bayesian philosophy of \citet{dosso_2006}.
\begin{table}

\centering
\footnotesize
\caption{Summary of inversion tests performed}
\label{alg:method}
\begin{tabular}{|p{\columnwidth}|}
\hline
 \hfil\textbf{Approach used for inversion} \\ \hline
    \textbf{1.} $k=0.03$ for all channels.\\
    \textbf{2.} $\textbf{k}$ as a vector.\\
    \textbf{3.} $\textbf{k}$ as a vector + $\mathbf{R}$ from HC repeat data with exponential roll-off.\\
    \textbf{4.} $\textbf{k}$ as a vector + $\mathbf{R}$ from residuals of \textbf{2.} with exponential roll-off.\\
    \hline
\end{tabular}
\end{table}
\section{Tests on real data from Menindee, 2024}
\label{sec:results}

\begin{figure*}
    \includegraphics[width= \textwidth]{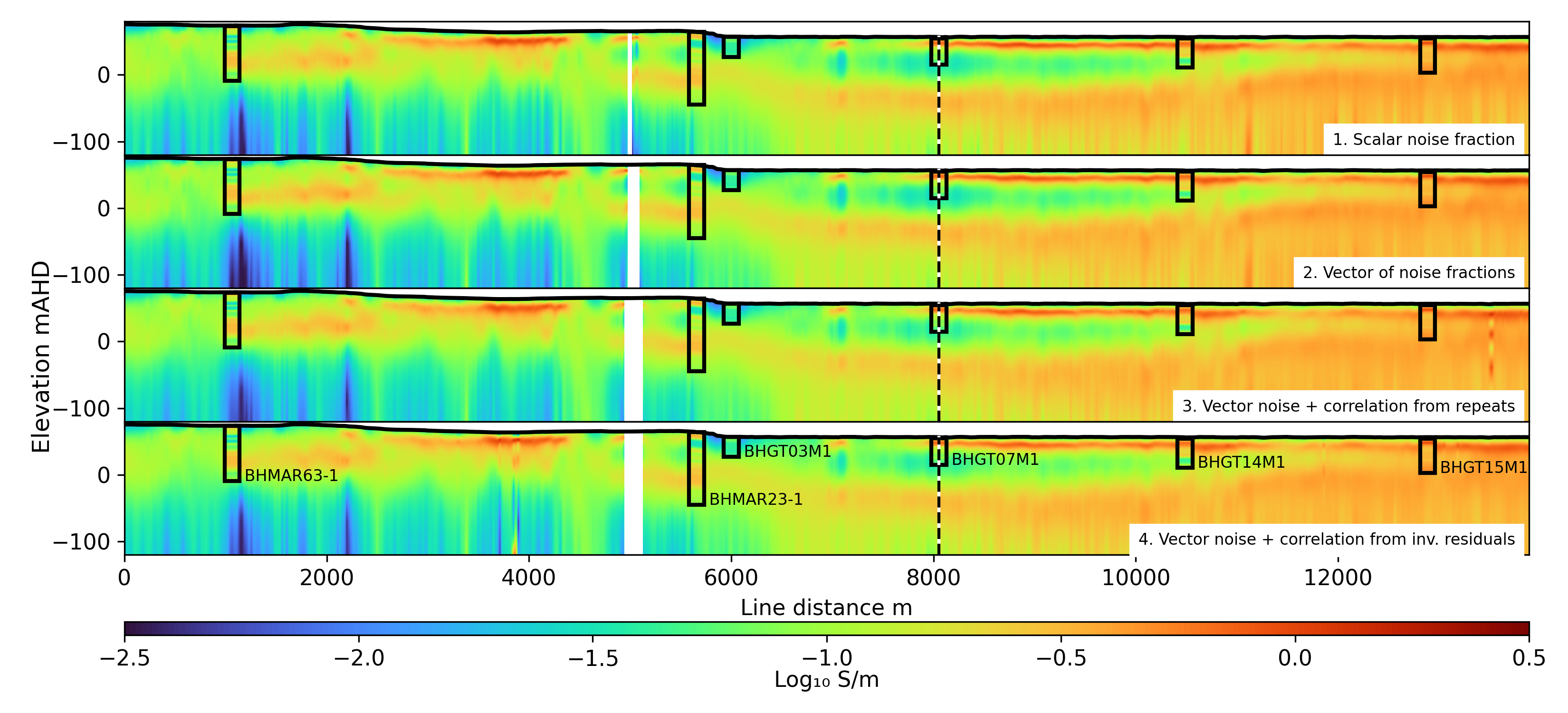}
    \caption{Cross-sections of inverted conductivity with depth of the same line across Menindee Lake from the approaches in Table~\ref{alg:method}. Borehole induction log conductivities are overlain using the same colour scale. Warmer colours are conductive and cooler colours are resistive. Areas that are blanked are where the $\phi_d$ misfit exceeds 2.25, i.e. the RMS misfit is greater than 1.5. The dashed profile through BHGT07M1 is examined in Section~\ref{sec:normalised_residuals_analysis}.}
    \label{fig:cross_sections}
\end{figure*}
\begin{figure}
    \centering
    \includegraphics[width=\columnwidth]{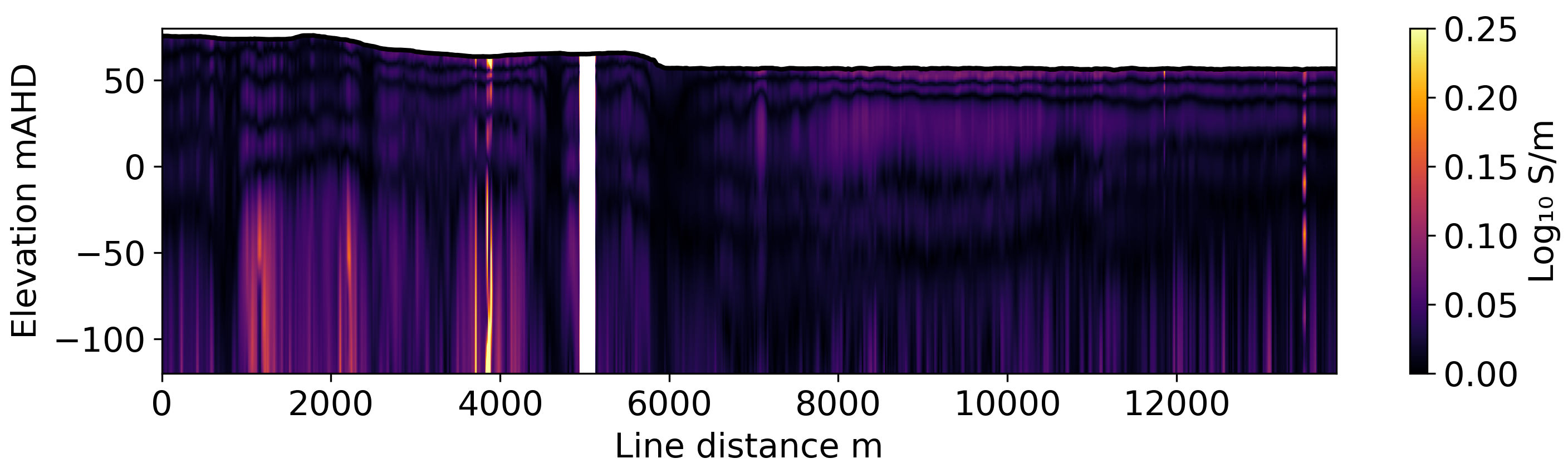}
    \caption{Pixel-wise standard deviation of inverted conductivity from all inversion tests in Figure~\ref{fig:cross_sections}. Lighter colours show higher standard deviation. Blanked areas show soundings where at least one test has $\phi_d$ exceeding 2.25 and we do not consider the inversion to have converged.}
    \label{fig:sd_comparisons}
\end{figure}

We carry out a series of real data inversion tests using the methodologies outlined in the previous section. These tests are summarised here briefly before we describe them in detail. 

In our first test we perform an inversion using a scalar multiplicative estimate of $k=0.03$. Recall that this is estimated using the non-linear height correction outlined in Section~\ref{sec:repeat_lines_estimating_method}. In the second test we use a time varying multiplicative noise factor as described in section \ref{sec:noise_vec}.

The third test implements correlation within the data covariance using the first of the two methods explained in Section~\ref{sec:iccorr}. Namely, we estimate the correlation matrix from repeat line data calculated from the height correction. We construct covariance $\mathbf {C_d}$ with the time-variable multiplicative noise $\mathbf{S}$ and correlation $\mathbf{R}$ from the HC data.  In the last test, we use correlation estimated from inversion residuals. These residuals are representative of the underlying noise process \citep{dosso_2006} and can mitigate the need for repeat lines. As in the previous test, we construct $\mathbf {C_d}$ with the time-variable multiplicative noise $\mathbf{S}$ but use $\mathbf{R}$ estimated from the inversion residuals of the second test.This results in four different approaches, summarised in Table \ref{alg:method}. Both the correlated case studies use a damped correlation as given by~\eqref{eq:exproll-off} and shown in Figure~\ref{fig:correlations}.

We clarify here, that $\mathbf d$ in~\eqref{eq:deterministic_misfit} is never height corrected. The HC data obtained from the repeat lines are only ever used to estimate noise parameters that feed into $\mathbf{C_d}$, which is used in~\eqref{eq:phid} through \eqref{eq:wtw}. The inverted data is never height corrected.

\subsection{On the validity of estimated correlations}
\label{sec:normalised_residuals_analysis}

In Section~\ref{sec:iccorr}, two methods were presented for estimating inter-channel noise correlation: from repeat line data, and from inversion residuals. Both approaches rest on the assumption that the residuals used are representative samples of the noise process. Here we examine this assumption more carefully in the context of regularised inversion.

As shown in Appendix~\ref{sec:pointest}, even for a linear problem with an Occam-type regularised solution, the residuals from this solution are biased, and the covariance of the residuals is not the data covariance matrix. Therefore, when one computes the sample covariance of standardised inversion residuals as in Section~\ref{sec:iccorr}, the result is not the true noise correlation $\mathbf R$, but a version of it filtered through the least squares machinery detailed in Equation~\eqref{eqn:rescov}, and potentially contaminated by the bias term in Equation~\eqref{eqn:resbias}.

A similar issue affects the repeat line approach, though in a different manner, and consequently to a lesser degree. As the repeat observations are acquired at different flight heights, a height correction is required before computing statistics (Section~\ref{sec:repeat_lines_estimating_method}). This correction depends on a forward prediction from a regularised model, and thus the regularisation bias enters the repeat line covariance estimate as well.

Therefore, neither method provides an uncontaminated estimate of $\mathbf R$. In practice, when both repeat line and inversion residual estimates of $\mathbf R$ are available, we observe that the raw correlation matrices differ, reflecting the different ways in which regularisation contaminates each estimate \ref{fig:correlations}(a,c).

\subsection{Comparison of inversion models using different noise estimation methodologies}
\label{sec:subsurface_results}
Carrying out the inversions in the sequence described in Table~\ref{alg:method}, we compare the results from each approach in each row of Figure \ref{fig:cross_sections}. We also provide approximate real-world validation, by overlaying borehole conductivity logs (with a much smaller footprint) in the same colour scheme as the inversion conductivities. The first striking observation is that despite differences in the covariance $\mathbf {C_d}$ used in \eqref{eq:phid}, whether with constant or variable $k_j$, with or without correlation, the differences between each row are relatively small. The match to the borehole conductivities is also qualitatively reasonable. 

Figure \ref{fig:sd_comparisons} displays the standard deviation per pixel, of inverted conductivity between all 10 tests. This highlights that the conductivity changes between different tests are minor, not exceeding a standard deviation of 0.25 $\log_{10}$ S/m outside the non-converged areas.  A detailed analysis of the inversion residuals in each approach, and a closer look at the inverted conductivity models through the dashed profile coincident with borehole BHTG07M1 in Figure~\ref{fig:cross_sections} is carried out in the next section.
\begin{figure*}
        \includegraphics[width=0.48\textwidth]{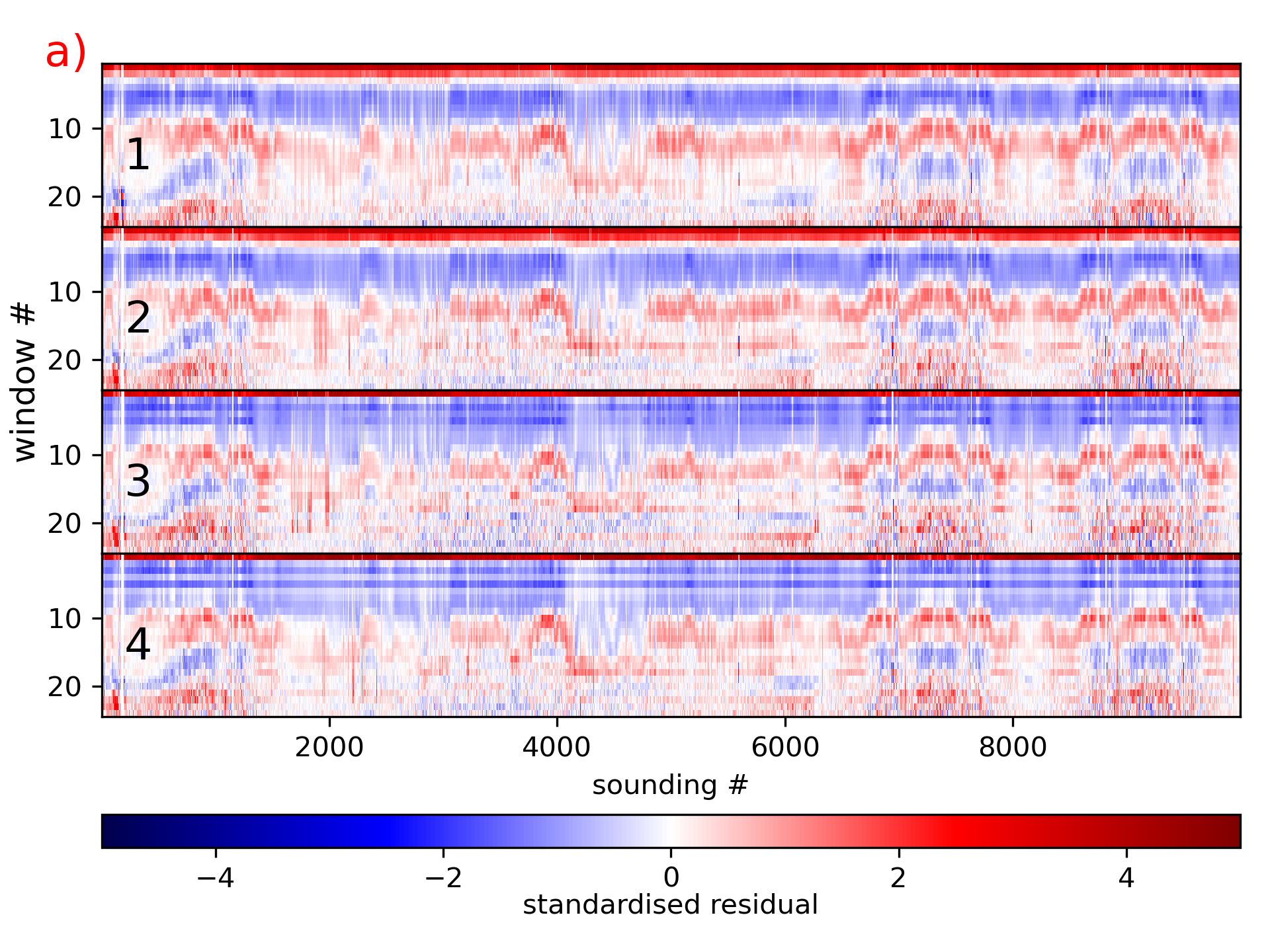}
        \label{fig:resids_and_differences}
        \includegraphics[width=0.48\textwidth]{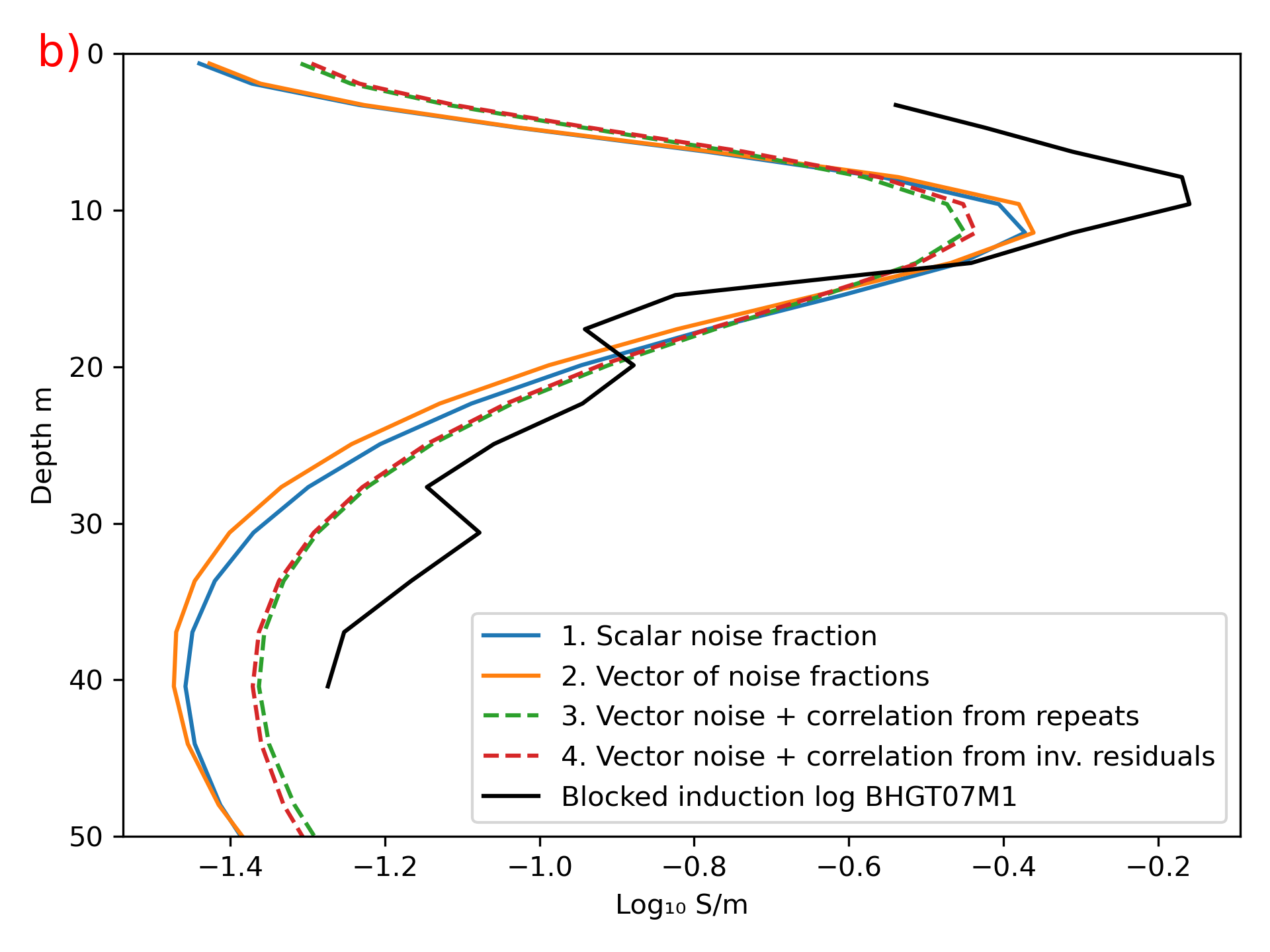}

        \caption{Residuals from all approaches (a) in Table~\ref{alg:method}, and profiles of conductivity (b) from each approach at  borehole BHGT07M1. The residuals in approaches 1 and 2 are from inversions that do not include correlation. The residuals from approaches 3 and 4 that do include time-window correlation, clearly show whitening, especially at early times -- note the thinner ``red band'' in the first channel and broken up nature of residuals in the first five channels when including correlation. As can be seen from the conductivity profiles, inclusion of correlation does lead to a difference in inverted models (dashed lines), though not significant enough to warrant any differences in geological interpretation. The borehole induction conductivity (black line) has been plotted in (b), and shows that models from all approaches are in good agreement with downhole values.}
        \label{fig:res_and_cross_section}

\end{figure*}
\begin{figure}
\centering
    \includegraphics[width= \columnwidth]{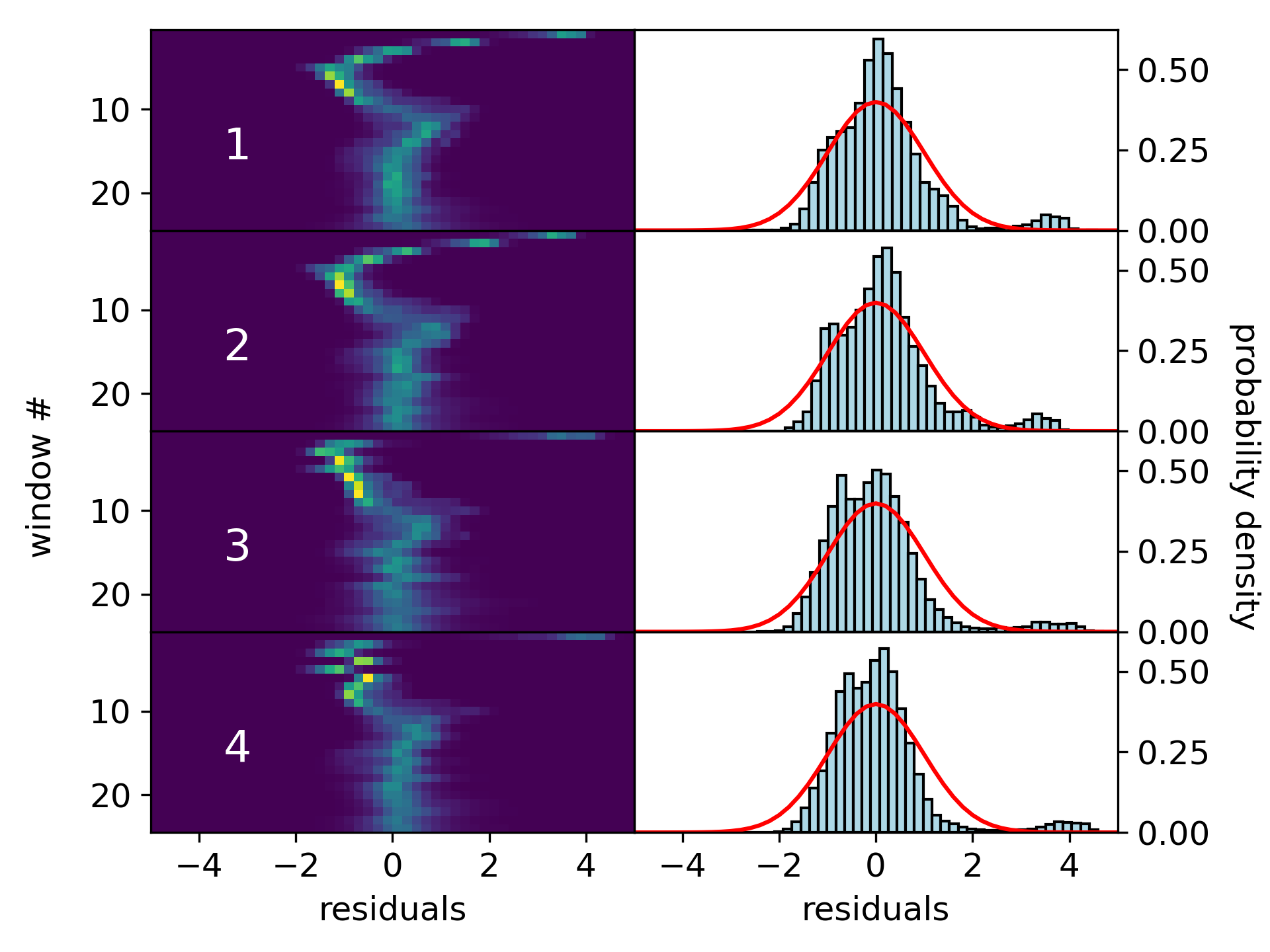}
    \caption{An analysis of standardised inversion residuals arranged from top-bottom according to Table~\ref{alg:method}. For each approach, the left column displays residuals for each time channel from all repeat lines. Colours that tend towards yellow indicate a more dense distribution of residuals. The right column displays the marginal distribution of the left column (i.e., aggregate over all times). The red line represents the $\mathcal{N}(0,1)$ probability density distribution.}
    \label{fig:gaussian_summary}
\end{figure}
\subsection{Analysis of inversion residuals and their statistics}
\label{sec:normalised_residuals_analysis}
 The standardised inversion residual for time channel $j$ at a particular sounding is given by $r_j = [\mathbf{U}^{-t}(\mathbf{d} - \mathbf{f}(\mathbf{m}))]_j$, specifically involving the whitening factor $\mathbf{U}^{-t}$. Examining residuals for all soundings, for every time channel, we can gain insights into which noise estimates, and hence inversion images, can be better trusted independent of borehole control. Figure \ref{fig:res_and_cross_section}a shows standardised residuals for every approach in Table~\ref{alg:method}. A smooth, continuous pattern through time represents correlation in the residuals. Better likelihood functions with appropriate noise characterisation should reveal that their pattern is more ``broken-up'' or de-correlated. Comparing diagonal covariance approaches (1 and 2), versus approaches which do use correlation (3 and 4), suggests improvement in the de-correlation of the residuals in approaches 3 and 4, especially in the early channels. These residuals when using a correlation matrix appear more ``high-frequent'' over time \citep[e.g.,][]{Dettmer2012}. 

In Figure \ref{fig:res_and_cross_section}b, examining closely the conductivity profiles from each approach at borehole BHGT07M1, we find that there is indeed a clear difference, even if minor from a geological interpretation point of view, as a consequence of including the correlation. As observed by other workers \citep[e.g.,][]{Dettmer2012a, https://doi.org/10.1111/1365-2478.70021}, inclusion of correlation usually leads to less structure in the inverted model. The effect is not as great as in probabilistic Bayesian inversions, due to the influence of regularisation in deterministic inverse problems as detailed in Appendix~\ref{sec:pointest}. A more direct way to understand this is by simplifying the objective function~\eqref{eq:deterministic_misfit} as the sum of the data misfit and regularisation terms:
\begin{align}
 \phi &= \mathbf{r}^t\mathbf{C_d}^{-1}\mathbf{r} +\lambda^2(\cdot),\\
 &= \mathbf{r}^t\mathbf{(SRS)}^{-1}\mathbf{r} +\lambda^2(\cdot),\\
 &= \mathbf{(S^{-1}r)}^t\mathbf{R}^{-1}\mathbf{S^{-1}r} +\lambda^2(\cdot), \label{eqn:split}
\end{align} 
where the data residual is written as $\mathbf r$, and the covariance is decomposed as in~\eqref{eq:Cd_decomposition} into a diagonal part with the standard deviations $\mathbf S$ and the correlation in $\mathbf R$. The $(\cdot)$ term contains the model norm. When considering two different inversions that use the same $\mathbf{S}$, and $\lambda^2$ is large enough to produce appreciable smoothing, unless $\mathbf R$ is appreciably different between two inversions, $\phi$ is not much different in the two cases. Consequently, the two model estimates must be similar from such inversions. Finally, since the damped correlations in Figure~\ref{fig:correlations}(b,d) are close to identity, this further weakens the effect of correlation and the models estimated with correlated data error are therefore not much different from those estimated without.

Figure \ref{fig:gaussian_summary} provides an overview of the results of each inversion (arranged from top to bottom according to Table~\ref{alg:method}) showing the distribution of the standardised residuals in each channel, as well as their marginal distribution across all channels. We would like for the marginal standardised residuals to closely follow a Gaussian distribution, thus validating data error statistics. These comparisons suggest that, except for the first two approaches that ignore correlation, all residuals are more ``broken-up" when including off-diagonal elements in \textbf{W}. It must be noted however, that though we have included a correlation matrix in the misfit function~\eqref{eq:phid}, as noted by \cite{tenorio2017introduction} and discussed in detail in Appendix~\ref{sec:pointest}, regularised least squares inverse solutions will always display some degree of bias and residual correlation.

We conclude that correlation estimated from either method, provides an approximation to the true noise correlation. While the theoretical analysis of Appendix~\ref{sec:pointest} shows that the estimate cannot be exact, it moves the noise model in the correct direction from the incorrect assumption of $\mathbf{R = I}$. If we have produced less structure in whitened residuals (Figure~\ref{fig:res_and_cross_section}a), the possibility of the ignored correlation entering the inversion estimates of conductivity and producing artefacts is lower, even if the whitened residuals can never be independent and identically distributed. We stress again, that from a geological interpretation point of view, the introduction of correlation into the deterministic AEM inverse problem is unwarranted due to the similarity of conductivity estimates obtained \citep[e.g.,][]{rayDecadeAirborneElectromagnetic2026}.

\section{Discussion and conclusions}
\label{sec:discussion}
In summary, we have provided details of a non-linear method used to estimate AEM multiplicative noise that dominates at early-to-mid times. Knowledge of this noise is crucial in performing an effective inversion to image subsurface conductivity. We have also provided a systematic methodology to sample the data noise correlation. 

Figure \ref{fig:gaussian_summary} demonstrates that some noise estimation approaches do demonstrate slight improvements of residuals across time channels over all lines. Channel-wise residuals are broken up to remove correlations that violate assumptions made about the data noise, especially at early times. 

There are a few important takeaways from our study on including correlation. Correlation estimated from inversion residuals will be a filtered version of the true correlation. An inverted model based height correction which allows correlation estimation from repeat lines is better. However, since deterministic inversion involves regularisation, bias enters the estimated correlation through the model based corrections themselves. Lastly, since there are far more time windows than repeat lines, for both the multiplicative noise as well as correlation estimates, we assume that the noise process is stationary over all survey flights. This is not unreasonable since all survey lines in an area are similarly flown, and all soundings are similarly filtered or processed. However, this could explain why inversions with undamped raw or lightly damped correlation do not converge, since estimated correlations may need to vary spatially. This could be another reason why the inverted conductivity models are not appreciably changed even after including correlation. To a large extent, these issues can be circumvented if Markov chain Monte Carlo (McMC) or similar Bayesian inversions that rigorously sample proportional to the likelihood in~\eqref{eq:gauss} are used. As shown in \cite{dosso_2006}, McMC inversion residuals can indeed closely follow the data noise statistics. An improved workflow could potentially make height corrections from the mean or median of McMC inversion models, or sample correlations from McMC inversion residuals as suggested by \cite{dosso_2006}. McMC inversions can also use hierarchical methods to estimate noise hyper-parameters \citep[e.g.,][]{bodin_2012}. Of course, McMC inversion is computationally expensive, but these are promising avenues for future work.

In standard deterministic inversion workflows, ensuring that residuals are decorrelated requires additional trial-and-error steps to adequately smooth the sampled correlation matrix. Our studies suggest that a non-linear height correction and estimating a constant $k$ or time-varying $k$, i.e., using a diagonal data covariance seems sufficient (Approach 1 in Table~\ref{alg:method}, as detailed in Section \ref{sec:repeat_lines_estimating_method}) -- this is an important finding for rigorous yet practical AEM inversion. 

This study was conducted using the NRG-XCITE, rotary-wing system.  It may still be useful to conduct these tests with other airborne systems and possibly additional testing sites, though we expect to draw similar conclusions.
\section{Acknowledgements}
This project was undertaken as part of the Geoscience Australia graduate program under the High Quality Geophysical Analysis (HiQGA) module, under the Resourcing Australia's Prosperity initiative (RAPi). HiQGA was first developed under the Exploring for the Future (2016-2024) program. This work used computing resources from the National Computational Infrastructure's (NCI) \textit{Gadi} supercomputer. The NCI is supported by the Australian Government. All code was written using the Julia language \citep{bezanson_2017}. We thank Roslynn King and an anonymous reviewer for comments that significantly improved this manuscript.
%%%%%%%%%%%%%%%%%%%%%%%%%%%%%%%%%%%%%%%%%%%%%%%%%%
\section*{Data Availability}
Test range data are provided in a commercial in confidence context and cannot be shared -- however, we use and modify the open-source HiQGA code base \citep{ray_2023,ray_2023_hiqga} to perform the geophysical inversions. The code base with example data can be freely cloned from \begin{verbatim}https://github.com/GeoscienceAustralia/HiQGA.jl\end{verbatim}

\section*{Conflict of Interest}
The authors declare no conflict of interest.



\bibliography{sample} % if your bibtex file is called example.bib

%\begin{thebibliography}{99}
%\end{thebibliography}

%%%%%%%%%%%%%%%%%%%%%%%%%%%%%%%%%%%%%%%%%%%%%%%%%%

%%%%%%%%%%%%%%%%% APPENDICES %%%%%%%%%%%%%%%%%%%%%
\appendix
\section{Optimising variance to find closest Gaussian to a given set of samples}
\label{sec:rationale}
\begin{figure}
\centering
    \includegraphics[width= \columnwidth]{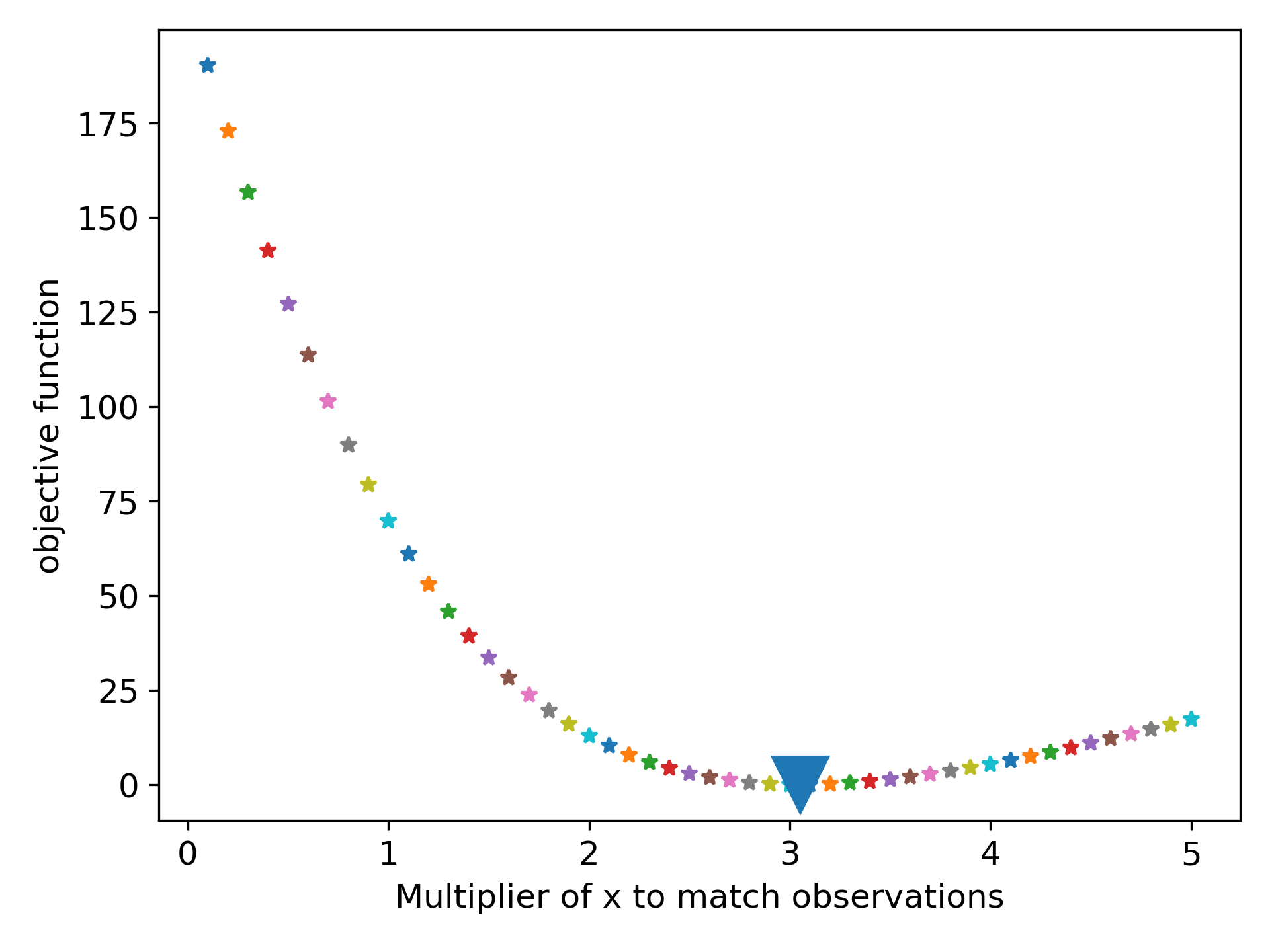}
    \caption{A line search over c in equation~\eqref{eq:noise_vec} for a synthetic set of 5,000 samples with variance $(\frac 1 3)^2$. The blue triangle shows the gradient descent result when starting from an initial guess of $c=5$. The true value is $c=3$.}
    \label{fig:linesearch}
\end{figure}
\begin{figure}
\centering
    \includegraphics[width= \columnwidth]{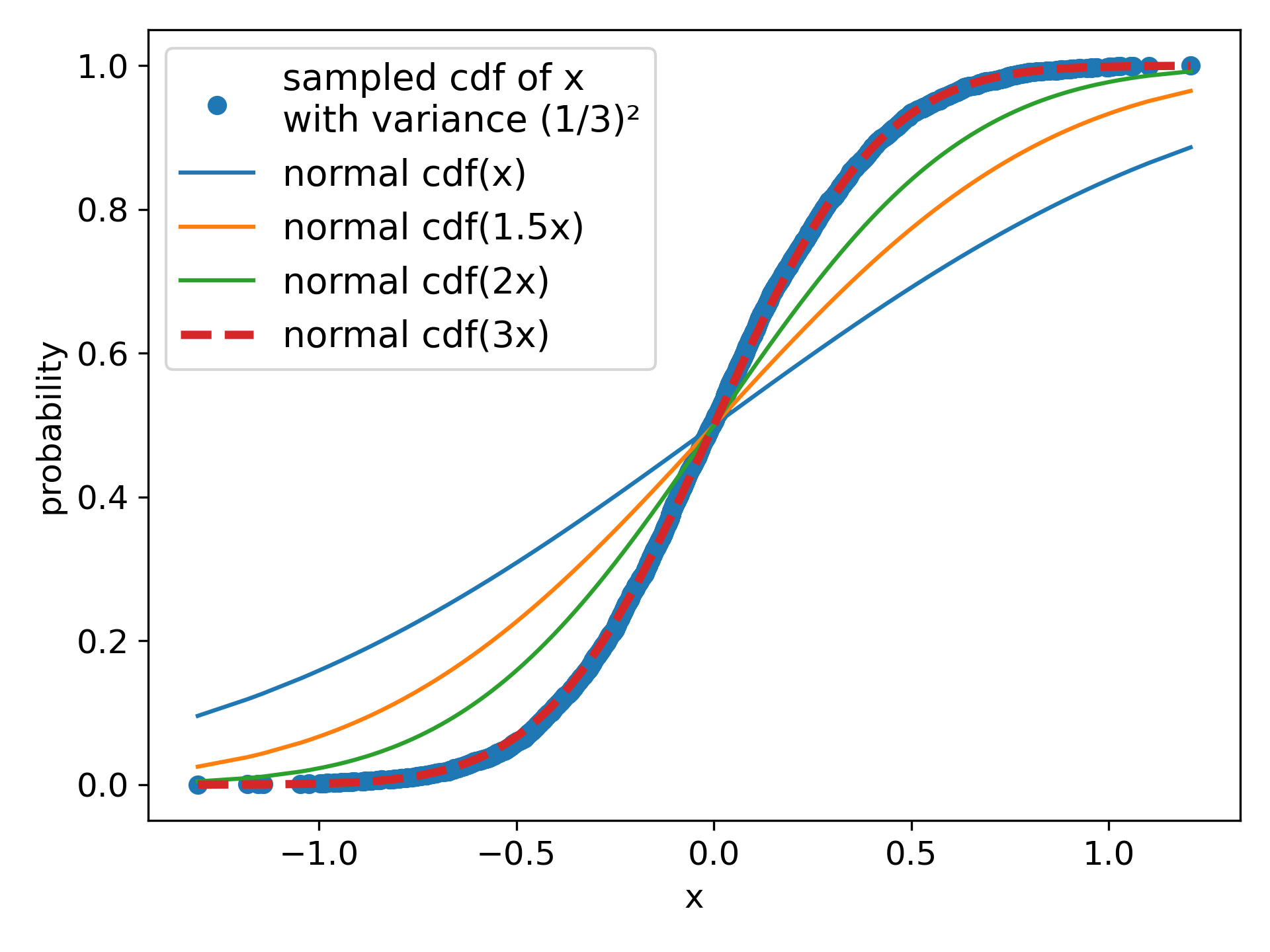}
    \caption{If we have a set of residuals $x$, akin to our height corrected AEM residuals divided by mean amplitude shown in Figure~\ref{fig:noise_vec}, $x$ needs to be multiplied by $c$, i.e., the mean amplitudes need to be multiplied by $1/c$ to match the CDF of the observations. A sweep through different values of $c$ is shown in this figure.}
    \label{fig:scaled_CDF}
\end{figure}
We use~\eqref{eq:noise_vec} to find the factor $c$, that when multiplied into the given samples and evaluated through a standard Gaussian cumulative density function, matches the observed samples. This is demonstrated here using a synthetic set of 5,000 normally distributed samples with (unknown) variance $(\frac 1 3)^2$. A logarithmically spaced line search over $c$ using ~\eqref{eq:noise_vec} is shown in Figure~\ref{fig:linesearch}. The estimate from a standard Broyden-Fanno-Goldfarb-Shanno gradient descent algorithm \citep{wright2006numerical} is shown with the blue triangle. As expected, the empirical CDF of $x\sim\mathcal N(0,(\frac 1 3)^2)$, matches CDF$_{\mathcal{N}(0,1)}(3x)$, as demonstrated in Figure~\ref{fig:scaled_CDF}. Essentially, we vary $c$ until the CDF for the unit Gaussian scales to the CDF of the observations with actual variance $(\frac 1 3)^2$ at $c=3$.

\section{Covariance of residuals for a regularised least squares problem}\label{sec:pointest}
For a simple linear problem of the form 
\begin{align}
\mathbf {y} &= \mathbf{Ax} + \bm\epsilon \;\text{where }\bm{\epsilon} \sim \mathcal{N}\mathbf{(0, C)}, \label{eq:linear}
\intertext{when regularising with an operator $\mathbf L$, an Occam-type regularised least squares solution can be found as}
\mathbf{\hat x} &= (\mathbf A ^t  \mathbf C^{-1}  \mathbf A + \lambda^2  \mathbf L^t  \mathbf L)^{-1} \mathbf A ^t \mathbf C^{-1}  \mathbf y.
\intertext{Writing the residual from the least squares estimate as:}
\mathbf{\hat r} &\equiv \mathbf {y -A\hat x},\\
&= \mathbf{y} -  \underbrace{\mathbf A (\mathbf A ^t  \mathbf C^{-1}  \mathbf A + \lambda^2  \mathbf L^t  \mathbf L)^{-1} \mathbf A ^t \mathbf C^{-1}}_{\mathbf H}  \mathbf y, \\
&= \mathbf{(I -H)y}, \\
\intertext{now using the true value of $\mathbf x$ from \eqref{eq:linear},}
\mathbf{\hat r} &= \mathbf{(I -H)}(\mathbf{Ax} + \bm\epsilon),\\
\mathbf{\hat r} &=  \underbrace{\mathbf{(I -H)}\mathbf{Ax}}_{\text{deterministic bias}}+ \;\;\mathbf{(I -H)}\bm\epsilon,  \label{eqn:resbias} \\
\intertext{with covariance}
\text{Cov}(\mathbf{\hat r}) &= \mathbf{(I -H)}\big<\bm\epsilon \bm\epsilon^t\big>\mathbf{(I -H)}^t,\\
&=\mathbf{(I -H)} \mathbf C \mathbf{(I -H)}^t, \text{which is not the same as $\mathbf C$.}\label{eqn:rescov}
\intertext{Whitening $\mathbf{\hat r}$ with the Cholesky factor $\mathbf U$ such that $\mathbf U^t \mathbf U = \mathbf C,$ we can write}
 \mathbf{\hat r_w} & = \mathbf U^{-t} \mathbf{\hat r} = \underbrace{\mathbf U^{-t} \mathbf{(I -H)}\mathbf{Ax}}_{\text{bias constant}} \,+ \,\mathbf U^{-t}\mathbf{(I -H)}\bm\epsilon. \label{eqn:whit}
\end{align}
Therefore, even when $\mathbf C$ is known, the whitened residuals are \textit{always} biased, and using only the second term on the right hand side of Equation~\eqref{eqn:whit}, we can calculate the covariance of $\mathbf{\hat r_w}$ as follows:
\begin{align}
\text{Cov} (\mathbf{\hat r_w}) &= \mathbf U^{-t}\mathbf{(I -H)}\big<\bm\epsilon \bm\epsilon ^t \big> \mathbf{(I -H)}^t \mathbf U^{-1},\\
&=\mathbf U^{-t}\mathbf{(I -H)}\mathbf C \mathbf{(I -H)}^t \mathbf U^{-1}. \label{eqn:whitenedcov}
\end{align}
The right hand side of Equation~\eqref{eqn:whitenedcov} is not the identity matrix (perfect whitening), unless $\mathbf H \approx \mathbf 0$, which can happen as $\lambda^2 \rightarrow \infty$. However, in this case the data are not being fit and residuals $\mathbf{\hat r}$ are the data. This reflects estimates from an ``exact'' repeat line scenario requiring no height correction, where we would not need inversion residuals to calculate correlation. Therefore $\lambda^2 \rightarrow \infty$ is only a theoretical comfort. At the opposite end, if $\lambda^2 \rightarrow 0$, then $\mathbf H$ is significantly large and the estimate of $\mathbf C$ is highly distorted. It is of some comfort that $0 \ll \lambda^2 \ll \infty$ in the linearised AEM inversion step, thus yielding estimates of correlation that are not massively distorted.

%%%%%%%%%%%%%%%%%%%%%%%%%%%%%%%%%%%%%%%%%%%%%%%%%%

% Don't change these lines

\label{lastpage}
\end{document}